\def\be{\begin{equation}}
\def\ee{\end{equation}}
\def\bea{\begin{eqnarray}}
\def\eea{\end{eqnarray}}
\def\noin{\noindent}
\def\V{{\bf{V}}}

\def\x{{\bf{x}}}

\def\k{{\bf{k}}}
\def\p{{\bf{p}}}
\def\q{{\bf{q}}}
\def\r{{\bf{r}}}

\def\Gammab{\overline{\Gamma}}
\def\intl{\int_0^{\Lambda}}

\documentstyle[aps,twocolumn,psfig,epsfig]{revtex}
\begin{document}
\draft
%\tighten
% The following line is crucial to get two-column format right
%\twocolumn[\hsize\textwidth\columnwidth\hsize\csname @twocolumnfalse\endcsname %\vskip2pc] \narrowtext
\title{Spectrum of  the Fokker-Planck operator 
representing diffusion in a random velocity field}
\author{ J.~T.~Chalker$^a$  and  Z.~Jane Wang$^{a,b\dagger}$}
\address{$^a$ Theoretical Physics, Oxford University,
1, Keble Road, Oxford, OX1 3NP, United Kingdom\\
$^b$ Courant Institute of Mathematical Sciences,
New York University       
251 Mercer St., New York, NY 10012  }
\date{\today}
\maketitle
\begin{abstract}
We study spectral properties of the Fokker-Planck operator 
that represents particles moving via a combination
of diffusion and advection in a time-independent random velocity
field,
presenting in detail work outlined elsewhere
[J. T.~Chalker and Z. J. ~Wang, Phys. Rev. Lett. {\bf 79}, 1797 (1997)].
We calculate analytically the ensemble-averaged one-particle Green 
function and the eigenvalue density for this Fokker-Planck operator,
using a  diagrammatic expansion developed
for resolvents of non-Hermitian random operators, 
together with a
mean-field approximation
(the self-consistent Born approximation)
which is well-controlled
in the weak-disorder regime for dimension $d>2$.
The eigenvalue density in the complex plane is non-zero within 
a wedge that encloses the negative real axis. Particle motion is 
diffusive at long times, but for short times
we find a novel time-dependence of the mean-square displacement,
$\langle r^2 \rangle \sim t^{2/d}$
in dimension $d>2$, associated with the imaginary parts of eigenvalues.
\end{abstract}

\pacs{PACS numbers:  05.10Gg, 05.60.-k, 47.55.Mh, 02.70.Hm}

\section{Introduction}
In this paper we study classical diffusion in the presence of
advection by a random, time-independent flow field, with an emphasis on spectral properties
of the corresponding Fokker-Planck operator.
Our motivation is two-fold. First, the mathematical problem
of calculating the properties of a random, non-Hermitian differential
operator is interesting in its own right, 
and, as far as we know, has received little attention until recently. 
Second, spectral decomposition is a natural approach for 
investigating the diffusion-advection problem. 
Amongst other things, the spectrum contains information about the time-dependence of the 
effective diffusivity, and our results in fact
reveal a short-time regime which appears not previously 
to have been discussed. 

Diffusion-advection problems can arise in a variety of physical
settings, including turbulent diffusion of tracer particles in 
geological systems, 
the temperature field in Rayleigh-B\'{e}nard convection,
and flow through porous media; an extensive review is 
provided in the article by Isichenko\cite{isi92}.  
The interplay between advection and diffusion may
either greatly enhance or strongly inhibit
the long-time transport of particles,
depending on the nature of the flow field.
Broadly speaking, compressible flows 
inhibit transport, since fluid sinks at which flow lines converge
will act as particle traps, while incompressible flows
transport particles more effectively at long times than diffusion alone.
In either case, the effective diffusivity provides 
a macroscopic measure of the combined consequences of molecular 
diffusion and of advection.

Diffusion-advection problems have long been a subject
of research in fluid dynamics. Early work dates back
to that of Taylor \cite{tay21} and of Richardson \cite{ric26},
who proposed the notion of effective diffusivity.  
Subsequently, Batchelor analyzed one-point and two-point correlations 
in homogeneous turbulence, relating the diffusion coefficient to 
the temporal correlation of the velocities \cite{bat49b}. 
Recently, much analytic progress has been made
in a special model considered by Kraichnan \cite{kra68}, in which the velocity
field is zero-mean Gaussian distributed and delta correlated in time, a
special case for which the closure problem is absent. 
In contrast, our interest in the following is in flow fields
that are time-independent.

The effect of various forms of time-independent spatial disorder
on classical diffusion has attracted considerable attention
from statistical physicists, especially following the proof by
Sinai that a type of random advection in one dimension results in
dramatically sub-diffusive motion at long times \cite{sin82}.
Several groups of authors \cite{fis84,aro84,kra85,dee95} have investigated
scaling in these systems, particularly within the framework
of renormalization group theory. Behavior is dependent on
dimensionality, and long-time motion is diffusive only above 
an upper critical dimension, which is two if the advecting
velocity field has only short-range correlations. While the focus of that work,
reviewed in Refs.\,\cite{ale81,bou90} has been long times and
systems at or below the upper critical dimension, we restrict ourselves in the following 
to systems above the upper critical dimension, and consider transient 
as well as long-time behaviour.

In addition to the study of diffusion-advection
problems as critical phenomena, many rigorous results, 
providing bounds on  the effective diffusivity and on the limiting 
behavior at long times, have been proved by mathematicians
using the methods of multiscale analysis and 
variational principles\cite{ave89,fan96,str79}.  

Against the background of this varied literature, it is 
perhaps suprising 
that little seems to have been known until recently
about spectral properties of the Fokker-Planck operator 
for random diffusion-advection problems,
even though the corresponding aspects of
random Schr\"odinger operators have 
been studied very extensively.
In fact, the spectrum of the Fokker-Planck
operator depends very much on the nature of the velocity field 
responsible for advection.
In general, this velocity field can be separated into an
incompressible (divergence-free) part, 
and a remainder, which can be expressed as the gradient of a scalar potential.
Pure potential flow is a special case of some importance, especially since 
in one dimension any velocity field can be written in terms of a scalar potential. 
In this special case
the Fokker-Planck operator is related by similarity transformation
to a Hermitian operator \cite{ris89}, and therefore has a purely real spectrum.
Some previous work has built on this transformation,
showing that  anomolous diffusion at long times in one dimension
is connected with a logarithmic
singularity of eigenvalue density of 
the Fokker-Planck operator \cite{tos88}, while other work
has been concerned  in two dimensions with the opposite limit 
of purely incompressible flow, analyzing the connection with 
the quantum random flux problem, and studying numerically the 
spatial decay of the Green function \cite{mil96}.

More generally, as we show, the non-Hermitian nature of the
Fokker-Planck operator is an obstacle that prevents one from
transferring 
in a straightforward way the techniques developed for random Schr\"odinger operators. 
For an unrestricted flow field, one expects eigenvalues of
the Fokker-Planck operator to occupy a finite area of the complex plane.
As a result, the corresponding Green function is non-analytic
throughout this area, and established perturbative approaches to calculating  disorder-averaged Green functions, which depend on analytic continuation, are inapplicable. This difficulty has been faced previously in the study of 
certain ensembles of non-Hermitian random matrices,
for which the eigenvalues are known to be distributed uniformly within a circle
in the complex plane \cite{gin65,gir85}.
For these ensembles, special techniques have been developed \cite{som88,leh91},
which do not generalise immediately to spatially extended problems such as the one we are concerned with. Recently, several groups independently 
\cite{fei97,efe97a,cha97,jan97a} have found that one can make progress by constructing, from the
non-Hermitian operator of interest,
a Hermitian operator with a $2 \times 2$ block structure,
and applying standard methods to this enlarged Hermitian operator.
We describe this approach below, and apply it
to the Fokker-Planck operator,
treating the advection
term as a perturbation to the diffusion term, a limit
is known in the fluid dynamics as the small P\'{e}clet  number regime.

The general method described here might be applied 
to variety of problems involving random non-Hermitian operators
in which behavior at weak disorder is of interest.
A number of such problems have attracted recent attention, including
asymmetric neural networks \cite{som88} and the statistical mechanics
of flux lines in superconductors with columnar disorder \cite{hat96},
and the Schr\"odinger equation for particles moving in a
random imaginary scalar potential \cite{sim98}. 
Recently, building on the formalism outlined
in Sec.\,II, useful links have been established between \cite{mud98}
problems of this kind and certain Hermitian localization problems.

The present paper provides a detailed account of work 
presented previously in outline elsewhere \cite{cha97}.
In particular, we describe for the first time calculations for a random flow
field with arbitrary relative strength to the compressible and incompressible
components. The remaining sections are organised as follows. In Sec.\,II, we define the problem and outline the Green function approach
used in our calculations.  In Sec.\,III we apply this approach 
to the Fokker-Planck operator for the diffusion-advection problem with
a random velocity field. We obtain the Green function for the 
Fokker-Planck operator, and its eigenvalue density in the complex plane.
We describe numerical tests of our results in Sec.\,IV,
and present some technical details in two appendices.

\section{Principles of calculation}
Particles that move by a combination of diffusion and
advection have a density, $n$, which is a function
of time, $t$, and position, $\r$, and
evolves according to the Fokker-Planck equation
\be
{\partial_t n} = {{\cal L}_{fp}}(n) \equiv D 
\nabla^2 n - {\nabla} \cdot ({\V} n),
\label{fp-eq}
\ee
\noindent 
where $D$ is the molecular
diffusivity, and $\V$ is the background velocity field,
which we take to be time-independent and random. 
For definiteness, we apply periodic boundary conditions to $n(\r,t)$
at the surface of a $d$-dimensional cube of volume $\Omega$, 
but treat in detail only the 
limit $\Omega \to \infty$. The Fokker-Planck operator,
${\cal L}_{fp}$, is a sum of two contributions: the
diffusion term,  ${\cal L}_{0}(n)= D \nabla^2 n$,
is Hermitian, while 
the advection term, ${\cal L}_{1}(n)= - {\nabla} \cdot ({\V} n)$,
is not, and our attention throughout this paper
is focused on their combined effects on 
the properties of ${\cal L}_{fp}$ as a {\it random,
non-Hermitian} operator.

Our choice for the probability distribution of the 
velocity field is intended to be the simplest:
we take it to be Gaussian with zero mean, and with correlations of $\V(\r)$
that are as short-range as possible. It is specified by the
variance 
\bea
\langle V_\alpha(\k)V_\beta(\k')\rangle
&=& \Gamma_1 ( \delta_{\alpha\beta} - \frac{k_\alpha
k_\beta} {k^2}) \delta(\k+\k')  \nonumber\\
&+& \Gamma_2 ( \frac{k_\alpha k_\beta} {k^2}) \delta(\k+\k'),
\label{eq-gamma}
\eea
where
\bea
\V(\k) &=& (2\pi)^{-d} \int d^d\r e^{-i\k\cdot\r}\,\V(\x), 
\eea
$\langle \ldots \rangle$ indicates an ensemble average,
and $\Gamma_1$ and $\Gamma_2$ represent the strengths of 
the incompressible and irrotational parts of the velocity field, 
respectively.
For the diffusion-advection problem to be well-defined,
it is necessary to impose a short-wavelength cut-off,
$\Lambda$, on the spectrum of velocity fluctuations,
so that Eq.\,(\ref{eq-gamma}) is supplanted by
$V_{\alpha}(\k)=0$ for $|\k| > \Lambda$.
A dimensionless measure of the disorder strength
involves a frequency scale, which we denote by $\omega$:
$\gamma_i(\omega) = (\Gamma_i/|\omega|^2)(|\omega|/D)^{d/2+1}$
is the ratio, raised to the power $d+2$,
of the distances travelled by a particle
due to advection and to diffusion, in time $\omega^{-1}$.
The fact that $\gamma(\omega)\propto |\omega|^{(d-2)/2}$
identifies $d=2$ as the upper critical dimension for the problem:
in dimension $d>2$, as $\omega \to 0$, $\gamma(\omega) \to 0$
and one expects the long-time behaviour simply to be diffusive,
while for $d<2$, $\gamma(\omega)$ diverges as $\omega \to 0$,
and long-time behaviour is dominated by advection. We restrict our 
discussion in this paper to the regime above the upper critical dimension.

Our calculations center on the disorder-averaged one-particle
Green function,
which in the position-space and time domain is
$\left<n(\r,t)\right>$, where $n(\r,t)$ satisfies Eq.\,(\ref{fp-eq})
with initial condition $n(\r,0)=\delta(\r)$.
We obtain this from its Laplace transform,
\be
g(\omega) = \langle \frac{1}{\omega - {\cal L}_{fp}} \rangle\,,
\ee 
in which $\omega$, in general complex, is the transform
variable conjugate to $t$. 
It is useful to display the spectral decomposition of both these
Green functions.
Let $\langle L_{\lambda}|$ and $|R_{\lambda} \rangle$
be the left and right eigenvectors of ${\cal L}_{fp}$
with eigenvalue $\lambda$. For finite system size, the spectrum is
discrete, the eigenvalues are non-degenerate with probability
one, and $\{\langle L_{\lambda}|\}$, $\{|R_{\lambda} \rangle\}$
are complete, biorthogonal bases. 
Introducing a plane-wave basis, 
$\langle \r|\p\rangle = \Omega^{-1/2} e^{i\p.\r}$, the average 
spectral density is defined as
\be
C(\p,\omega) = \langle \sum_{\lambda}
\langle \p |R_{\lambda} \rangle \langle L_{\lambda} | \p \rangle
\delta(\omega - \lambda)\rangle\,.
\ee
Then
\be
\left<n(\r,t)\right> = (2\pi)^{-d}\int d^{d}\p e^{i\p.\r}
\int d^2\omega e^{\omega t}
C(\p,\omega)\,.
\label{eq-n}
\ee
The average Green functions are diagonal in the
plane-wave basis, and we denote the diagonal elements
of $g(\omega)$ by
$g^p(\omega)$:
\be
g^p(\omega) = \int d^2 \lambda \frac{C(\p,\lambda)}{\omega - \lambda}\,,
\ee
where the dependence is only on $p$, the magnitude of $\p$,
in the large $\Omega$ limit, because of spatial isoptropy.
Finally, the eigenvalue density is
\be
\rho(\omega)=\Omega^{-1}\sum_{\lambda}\delta(\omega-\lambda)\,.
\ee
Both $C(\p,\lambda)$ and $\rho(\lambda)$ can be calculated from the Green function using the
identity
\be
\delta(\omega-\lambda)=\frac{1}{\pi} \frac{\partial}{\partial \omega^*} \frac{1}{\omega-\lambda}
\label{eq-rho}
\ee
with the notation 
$\partial / \partial \omega^* = 
(1/2)(\partial /\partial x +i \partial/\partial y)$
for $\omega = x + i y$ and $x,y$ real.
The eigenvalue density
will turn out to be non-zero over a finite area of the complex
$\omega$ plane, and not just on the real axis as would be the case
if ${\cal L}_{fp}$ were Hermitian.

Standard methods \cite{agd} for calculating average Green functions for
random Hermitian operators make extensive use of the fact that they are
analytic in the complex $\omega$-plane, except on the real axis. By contrast,
$g(\omega)$ is non-analytic throughout the (as yet unknown) area of the 
complex $\omega$ plane in which $\rho(\omega)$ is non-zero.
Because of this, a new approach is required.
We first embed the original non-Hermitian  operator,
in the combination $\omega -{\cal L} \equiv {\cal A}$,
in a Hermitian operator ${\cal H}$ which has twice the 
dimension of ${\cal L}$, setting
\be
{\cal H} = {\cal H}_0 + {\cal H}_1 \ ,
\ee
where \cite{footnote1}
\be
{\cal H}_0 =\left( 
\begin{array}{cc}
\eta \hspace{0.5cm} 0\\
0 \hspace{0.5cm} -\eta\\
\end{array}
\right) \hspace{0.2cm} \mbox{and} \hspace{0.2cm} 
{\cal H}_1 =\left( 
\begin{array}{cc}
0 \hspace{1.0cm} {\cal A}\\
 {\cal A}^{\dagger} \hspace{1.0cm} 0\\
\end{array}
\right). 
\ee
The inverse of ${\cal H}$ exists for real $\eta \not=0$. Its average is
\bea
G &=& \left<{\cal H}^{-1}\right>  \equiv \left( 
\begin{array}{c}
G_{11} \hspace{0.5cm} G_{12}\\
G_{21} \hspace{0.5cm} G_{22}\\
\end{array}
\right) \nonumber  \\
&=&
\left( 
\begin{array}{c}
\langle\eta[\eta^2+{\cal A}{\cal A}^{\dagger}]^{-1}\rangle 
\hspace{0.5cm} \langle {\cal A}[\eta^2 +{\cal A}^{\dagger}{\cal A}]^{-1}\rangle\\
\langle{\cal A}^{\dagger}[\eta^2 +{\cal A}{\cal A}^{\dagger}]^{-1}\rangle 
\hspace{0.5cm}  
\langle -\eta[\eta^2+{\cal A}^{\dagger}{\cal A}]^{-1}\rangle\\
\end{array}
\right).
\label{eq-Gdef}
\eea
Note that the desired Green function is obtained using
$g(\omega) = \lim_{\eta \rightarrow 0}
G_{21}$. Since ${\cal H}$ is Hermitian, we can calculate $G$, and
hence $g(\omega)$, with extablished techniques.
Specifically,
$G$ can be expanded as a power series in ${\cal H}_1$ 
and in $G_0 \equiv {\cal H}_0^{-1}$
\bea
G &=& G_0 -\left<G_0 {\cal H}_1 G_0\right> + 
\left<G_0 {\cal H}_1 G_0 {\cal H}_1 G_0 \right> + .... 
\eea
which leads to a Dyson equation with self-energy $\Sigma$
\bea
G=G_0 + G_0 \Sigma G\,.
\eea

\begin{figure}[htb]
\centerline{\epsfig{figure=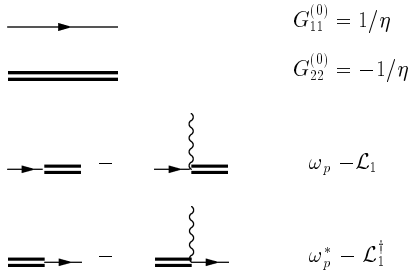,width=0.3\textwidth}}
\caption{Propagators and vertices.}
\label{fig-ver}
\end{figure}

Working in the plane-wave basis,
$G$ and $\Sigma$ each consist of four diagonal blocks, 
with elements 
$G_{ij}(p)$ and
$\Sigma_{ij}(p)$, for $i,j=1$ or $2$.  
It is convenient to represent the power series
diagrammatically. As shown in Fig.\,(\ref{fig-ver}),
we have two propagators, $G^{(0)}_{11} \equiv 1/\eta$ 
and  $G^{(0)}_{22} \equiv -1/\eta$, denoted by a single and double line,
respectively, and two vertices, ${\cal A}$ and 
${\cal A}^{\dagger}$. It is natural to
separate each vertex into a constant part and a random part:
case of $\cal A$, $\omega -{\cal L}_0$ and $-{\cal L}_1$.
$\omega - {\cal L}_0$ is diagonal, with diagonal elements
$\omega_p \equiv \omega + Dp^2$, and averages of the random parts
are specified by the four cumulants:
$\left<{\cal L}_1{\cal L}_1\right>$, 
$\left<{\cal L}_1{\cal L}_1^{\dagger}\right>$,
$\left<{\cal L}_1^{\dagger}{\cal L}_1\right>$, and
$\left<{\cal L}_1^{\dagger}{\cal L}_1^{\dagger}\right>$.

The leading contributions to $G$ at weak disorder
are resummed by the self-consistent Born approximation (SCBA).
This approximation is defined by the
expression for the self-energy shown diagrammatically 
in Fig.\,(\ref{fig-sigma}), in which the internal propagators
represent the full Green function $G$.  We demonstrate
in appendix A that corrections to the SCBA self-energy
are small in powers of $\gamma(|\omega|)$.

\begin{figure}[htb]
\centerline{\epsfig{figure=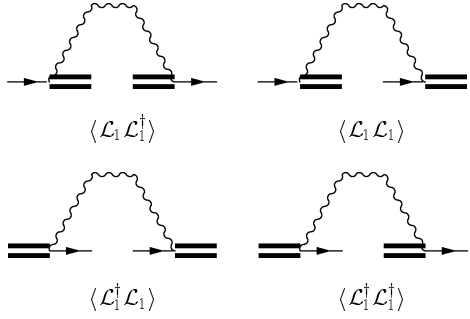,width=0.3\textwidth}}
\caption{Diagrammatic representation of disorder average.}
\label{fig-disave}
\end{figure}

\begin{figure}[htb]
\centerline{\epsfig{figure=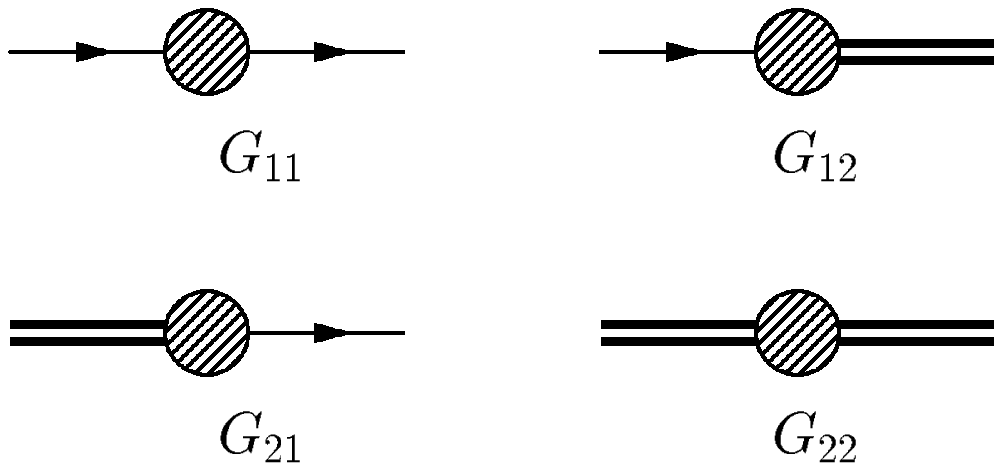,width=0.3\textwidth}}
\centerline{\epsfig{figure=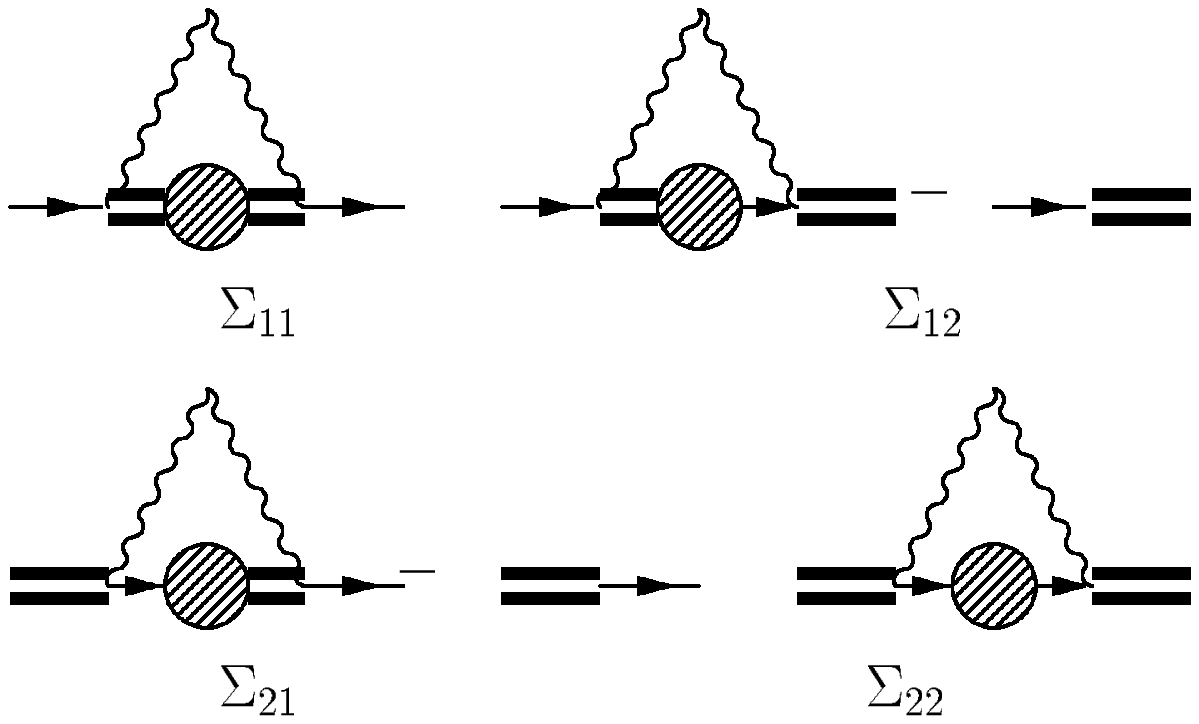,width=0.3\textwidth}}
\caption{Full Green function and  self-energy diagrams in the SCBA.}
\label{fig-sigma}
\end{figure}

There are useful interrelations among the four blocks of
$G_{ij}$, and also among those of $\Sigma_{ij}$.
Quite generally, in the limit $\epsilon \rightarrow 0$, 
one has $G_{21} = G^{\dagger}_{12}$ and $\Sigma_{21} = 
\Sigma^{\dagger}_{12}$; using the fact 
that each is diagonal in the plane-wave basis, one also has
$\Sigma_{11}(p) =-\Sigma^{*}_{22}(p)$
and $G_{11}(p) = - G^{*}_{22}(p)$.
It is therefore sufficient to
calculate only $G_{11}(p)$, $G_{21}(p)$, $\Sigma_{12}(p)$ and
$\Sigma_{22}(p)$. They are related via the Dyson equation,
which yields

\bea
G_{11}(p)&=&
[\eta+\Sigma_{22}(p)]C(p) \\
G_{21}(p)&=&
[-\Sigma ^{*}_{12}(p)]C(p)
\eea

where
\be
C(p) =[|\Sigma_{12}(p)|^2 +  |\eta+\Sigma_{22}(p)|^2]^{-1}\,.
\label{eq-c}
\ee

In turn, the SCBA generates expressions for $\Sigma_{ij}(p)$
in terms of $G_{ij}(p)$: referring to Fig.\,(\ref{fig-sigma}), we have
\bea
\Sigma_{12}(p) &=& -\omega_p
      +\intl d^dq \left<{\cal L}_1(p,q){\cal L}_1(q,p)\right> G_{21}(q)
\label{eq-s12}\\
      \Sigma_{22}(p) &=& 
      \intl d^dq \left<{\cal L}_1^{\dagger}(p,q)
     {\cal L}_1(q,p)\right> G_{11}(q)\,.
\label{eq-s22}
\eea
The solution to this set of equations, with the velocity correlations
of Eq.\,(\ref{eq-gamma}), simplifies considerably in the
special case $\Gamma_1 = \Gamma_2$, for which the calculations have been described previously \cite{cha97}. The solution in the general case is
presented in the following section.

\section{Application of SCBA}

The expressions for the self energy within the SCBA, Eqs.\,(\ref{eq-s12}) and (\ref{eq-s22})
become, with the Gaussian velocity distribution of Eq.\,(\ref{eq-gamma}),
\bea
\Sigma_{12}(p) = -\omega_p+ \Gamma_1 
                 \intl d^dq (\p \cdot \q)  
                 \Sigma^{*}_{12}(q) C(q) + \nonumber      \\
(\Gamma_2-\Gamma_1)
                 \intl \!\! \! \! \! d^dq \frac{[\q \cdot (\q-\p)][\p \cdot (\q-\p)]}
                 {|\q-\p|^2} \Sigma^{*}_{12}(q)
                 C(q)                    \label{eq-sig1} \nonumber      \\
\eea
\bea
\Sigma_{22}(p) = \Gamma_1
               \intl d^dq  q^2 [\eta+\Sigma_{22}(q)]C(q) +\nonumber \\
               (\Gamma_2-\Gamma_1)
               \intl d^dq \frac{[\q \cdot (\q-\p)]^2}{|\q-\p|^2} 
               [\eta+\Sigma_{22}(q)] C(q)       \label{eq-sig2}
\eea
where the fact that $\intl d^dq (\p.\q) \Sigma^{*}_{12}(q) C(q) = 0$
for symmetry reasons is an immediate simplification.
Eqs.\,(\ref{eq-sig1}) and (\ref{eq-sig2}), with Eq.\,(\ref{eq-c}),
are coupled integral equations, and presumably 
in general not exactly solvable. To make progress, we treat weak disorder,
for which the problem can be reduced to a system of algebraic equations.
As a starting point, consider the form of the function $C(q)$, Eq.\,(\ref{eq-c}), which appears in the integrands of 
Eqs.\,(\ref{eq-sig1}) and (\ref{eq-sig2}). First, in the {\it absence}
of disorder, $\Sigma_{12}(q) = -\omega_q$, $\Sigma_{22}=0$ and for $\eta \to 0$
\be
C(q) = \frac{1}{|\omega_q|^2} \equiv \frac{1}{(x +Dq^2)^2 + y^2}\,,
\ee
and so, if $\omega\equiv x+iy$ lies on the negative real axis,  $C(q)$ has a divergence as a function of $q$, at $q=q_0$, where $x+Dq_0^2=0$.
In the presence of weak disorder, we anticipate that
$C(q)$ will be large only for $x<0$, $|y|$ sufficiently small and 
$q\approx q_0$. We also expect that, if $\omega$ lies in the right half of the
complex plane, or if $|y|$ is large, so that $C(q)$ is small throughout the range of integration over $q$, then a good approximation for $\Sigma_{ij}(p)$ should be obtained from evaluating Eqs.\,(\ref{eq-sig1}) and (\ref{eq-sig2})
by iteration, using the disorder-free form for $C(q)$ in the integrands.
This approach is analogous to the first-order Born approximation of scattering theory.
Conversely, if $x<0$ and $|y|$ is small, we expect that it will be necessary
to determine $C(q)$ self-consistently, but with the simplification
at weak disorder that the dominant contribution to 
the integrals of Eqs.\,(\ref{eq-sig1}) and (\ref{eq-sig2})
is from the shell of wavevectors on which $q \approx q_0$.
For $x<0$, we therefore evaluate the integrals by replacing $C(q)$
with $I (\omega)\delta(q^2 - q_0^2)$, where $ I(\omega) = \intl d(q'^2) C(q')$.
This procedure yields the leading behaviour at small $\Gamma_1,\Gamma_2$,
provided the integrals involved are convergent at small and large $q$,
which is the case in dimensions $2<d<4$.
With this approach we find
\bea
\Sigma_{12}(p) &=& -\omega_p +(\Gamma_2-\Gamma_1)\frac{S_d q_0^d}{2}
\Sigma_{12}^{*}(q_0)I(\omega)\times \nonumber \\ 
& &\langle \frac{[\hat{n}\cdot(\hat{n}-\hat{n_0})]
                   [\hat{n_0}\cdot(\hat{n}-\hat{n_0})]}
                          {(\hat{n} -\hat{n_0})^2} \rangle_{\hat{n}}
\label{eqq-sig1}\\
\Sigma_{22}(p) &=& \frac{S_d q_0^d}{2 }
                [\eta+\Sigma_{22}(q_0)]I(\omega)\times \nonumber \\
& &\left[ \Gamma_1 + (\Gamma_2-\Gamma_1)
   \langle         \frac{[\hat{n}\cdot(\hat{n}-\hat{n_0})]^2}
                          {(\hat{n} -\hat{n_0})^2}  \rangle_{\hat{n}}                                  \right]\label{eqq-sig2}
\eea
where $S_d$ is the surface area of a unit sphere in $d$ dimensions,  
$\hat{n}$ and $\hat{n_0}$ are $d$-dimensional unit vectors
in directions corresponding to those of $\q$ and $\p$ in Eqs.\,(\ref{eq-sig1}) and (\ref{eq-sig2}), and $\langle \ldots  \rangle_{\hat{n}}$ denotes an 
angular average on ${\hat n}$. These averages have the values
\bea
\langle \frac{[\hat{n}\cdot(\hat{n}-\hat{n_0})]
                   [\hat{n_0}\cdot(\hat{n}-\hat{n_0})]}
                        {(\hat{n} -\hat{n_0})^2} \rangle_{\hat{n}} =
-\frac{1}{2}\,, \\
\langle \frac{[\hat{n}\cdot(\hat{n_0}-\hat{n})]^2}
               {(\hat{n} -\hat{n_0})^2} \rangle_{\hat{n}} 
= \frac{1}{2}\,.
\eea  
Notation is simplified by introducing the new variables:
$\Gamma_i^{'} = \Gamma_i S_d q_0^d/2$, 
$\Delta = (\Gamma_2^{'} - \Gamma_1^{'})/2$ and 
$\overline{\Gamma} = (\Gamma_1^{'} + \Gamma_2^{'})/2$.
In these terms, the solution to Eqs.\,(\ref{eqq-sig1}) and (\ref{eqq-sig2}) is
\bea
\Sigma_{12}(p) &=& -\left(x+Dp^2
                     +i\frac{y}{1- \Delta I(\omega)}  \right)\,. \label{eq-ss12}\\
\Sigma_{22}(p) &=& \frac{\eta \Gammab I(\omega)}{1-\Gammab I(\omega)} \label{eq-ss22}
\eea
Substituting these expressions
into Eq.\,(\ref{eq-c}),  we obtain
\bea
C(q) &=&[\left( x+Dq^2 \right)^2 +\beta^2]^{-1}\,,
\eea
where
\be
\beta^2 = \left(\frac{y}{1-\Delta I(\omega)}\right)^2 +
    \left(\frac{\eta}{1-\Gammab I(\omega)}\right)^2\,,
\ee
and hence (for $x \ll - \beta$)
\bea
I(\omega) \equiv \int^{\Lambda^2}_{0}d(q^2)C(q) \approx 
\int^{\infty}_{-\infty}d(q^2)C(q)=
\frac{\pi}{\beta D}\,.  \label{eq-I2}
\eea
Thus, finally, $I(\omega)$ satisfies 
\be
\left(\frac{y I(\omega)}{1-\Delta I(\omega)}\right)^2 +
\left(\frac{\eta I(\omega)}{1-\Gammab I(\omega)}\right)^2
=\left(\frac{\pi}{D}\right)^2\,.
\label{eq-I}
\ee

The behavior of the solution to this equation
in the limit $\eta \to 0$ depends on the value of $y$.
If $y$ is sufficiently large ($y > y_B$, where $y_B$ is determined
below),
$\Gammab I(\omega) < 1$ and
the second term on the left of Eq.,(\ref{eq-I})
vanishes as $\eta \to 0$.
The limiting solution is then simply
\be
I(\omega)=\frac{\pi}{Dy+\pi\Delta}\,.
\ee
In that case
\bea
g^p(\omega) &=& \lim_{\eta \rightarrow 0} G_{21}
=\left[Dp^2 + x + i(y +\Delta\pi/D)\right]^{-1}\,.\\
\eea
This expression is manifestly analytic in $\omega$, implying from Eq.\,(\ref{eq-rho}) that $\rho(\omega)$, the eigenvalue density of the Fokker-Planck operator, is zero in this part of the $\omega$-plane.
A similar analysis applies for $y < -y_B$.
In contrast, for $|y| < y_B$, the second term on the left of Eq.\,(\ref{eq-I})
remains non-zero as $\eta \to 0$ and the limiting solution is
\be
I(\omega) = \Gammab^{-1} \,.
\ee
The boundary between these regimes is evidently $|y|=y_B$ with
\bea
y_B = \frac{\pi \Gamma_1' }{D} = \frac{\pi S_d}{2} \gamma_1(x)\cdot|x| \propto \Gamma_1|x|^{d/2}\,.
\label{eq-bdy}
\eea
Inside this boundary,
\be
g^p(\omega)=\frac{Dp^2+x-iy\Gammab/\Gamma_1^{'}}{(Dp^2+x)^2+(\pi \Gammab/D)^2}\,,
\ee
which, in the same spirit as our representation of $C(p)$, may be written as
\be 
g^p(\omega)= -\frac{iy}{\Gamma_1^{'}}\delta(p^2-q_0^2)\,.
\ee
From this we obtain, for $x<0$ and $|y|<y_B$
\bea
\rho(\omega) =  \frac{D}{(2\pi)^{d+1}\Gamma_1|x|}.
\eea

Finally we turn to the time-evolution of the particle density.
From $g^p(\omega)$ we calculate
\be
C(\p,\omega) = \frac{1}{2\pi \Gamma^{'}_1} \delta(p^2-q_0^2)\,.
\ee
Using Eq.\,(\ref{eq-n}), we find 
\be
\langle n(\r,t)\rangle= \frac{1}{(2\pi)^d}\int d^d\p\, e^{i\p.\r}\,
e^{-Dp^2t}\,[\frac{\sin(\omega_0t)}{\omega_0t}]\,, 
\label{n}
\ee
where $\omega_0 = y_B$, evaluated at $x=-Dp^2$: 
$\omega_0 =\pi S_d\Gamma_1 p^d/2D$. 
The integrand has two factors:
the first one, $e^{-Dp^2t}$, would arise simply from diffusion;
the second,  $\sin(\omega_0t)/\omega_0t$, is a consequence of advection. 
At weak disorder, when the 
SCBA is a controlled approximation,
the advective factor differs significantly from $1$ only
where the diffusive factor is small, so that $\langle r^2 \rangle \sim Dt$,
indicating the normal diffusive behavior.
By contrast, at strong disorder (when the SCBA is simply
a mean-field approximation), it is the  advective factor 
that sets the width of the density profile at short times,
and  $\langle r^2 \rangle \sim (\Gamma_1 t/ D)^{2/d}$. 

\section{Comparison with numerical simulations}

As a test of our analytical approach, we compare results for the boundary
of the support of the  eigenvalue density with 
numerical calculations of the eigenvalue distribution. The numerical 
calculations involve diagonalising a discretised version of the Fokker-Planck operator, as set out in this section. In order that the test of analytical reults is as direct as possible, we adapt the analytical theory for the discretised operator, in the way summarised in Appendix B.
We can make the comparison for a two-dimensional system despite the fact
that the theory developed in Sec.\,III has long-wavelength divergences
in dimensions $d\leq 2$, since the system sizes we study are small enough to cut off the divergences.

\begin{figure}[htb]
\centerline{\epsfig{figure=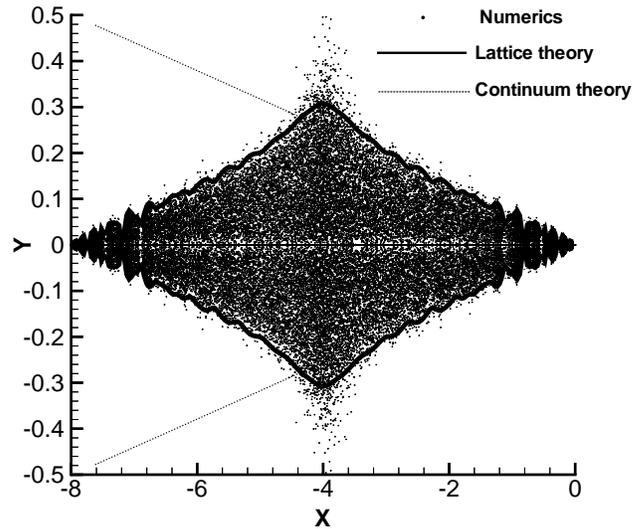,width=0.5\textwidth}}
\caption{ Distribution of eigenvalues in the complex plane (dots)
and calculated boundary to $\rho(\omega)$ for the lattice theory (full line)
and continuum theory (dotted line). }
\label{fig-egv}
\end{figure}

We discretise the Fokker-Planck operator keeping in mind the conservative form of the Fokker-Planck equation
\bea
\frac{\partial n}{\partial t} &=& \nabla \cdot J \\
J &=& D \nabla n - \V n\,.
\eea
On a two-dimensional, square lattice with coordinates ${i,j}$, velocity components
$u_{i,j}$ and $v_{i.j}$ are defined
on horizontal and vertical links, respectively, 
and densities $n_{i,j}$ at nodes. 
Using center difference, the discrete eigenvalue
equation becomes
\bea
\lambda\,n_{ij} = [D-\frac{1}{2}u_{i+\frac{1}{2},j}]n_{i+1,j} 
               + [D+\frac{1}{2}u_{i-\frac{1}{2},j}]n_{i-1,j} \nonumber \\
              + [D-\frac{1}{2}v_{i,j+\frac{1}{2}}]n_{i,j+1} 
                + [D+\frac{1}{2}v_{i,j-\frac{1}{2}}]n_{i,j-1}\nonumber \\
           + [-4D -\frac{1}{2}
                  (u_{i+\frac{1}{2},j} - u_{i-\frac{1}{2},j}
                  +v_{i,j+\frac{1}{2}} - v_{i,j-\frac{1}{2}})]n_{i,j}
\eea
We restrict ourselves to the case $\Gamma_1 = \Gamma_2 \equiv \Gamma$,
for which
the velocities $(u,v)$ are Gaussian random variables with
zero mean and variance $(2\pi)^2\Gamma$.  A cutoff $|u|\leq 2D$ is imposed 
to ensure that matrix elements of the discretized operator are non-negative, a requirement for the transition matrix in any Markov process, which in turn implies
that the real parts of the eigenvalues are necessarily non-positive.
The eigenvalues found by numerical diagnalization,
using 50 realisations of a $32 \times 32$ lattice 
with $D=1$ and $\langle u^2 \rangle = \langle v^2 \rangle = 0.25$
so that $\Gamma=0.25(2 \pi)^{-2}$, are shown in 
Fig.\,(\ref{fig-egv}), as well as
the boundary to the eigenvalue density
calculated for the discretized problem on a lattice of the same size (full line), 
and for the continuum theory (dotted line).
The full line is obtained, following Appendix B, as the
$\omega$ values which solve Eq.\,\ref{eq-dis}.
The dashed line is calculated from Eq.\,\ref{eq-con}: it has slope
$\Gamma\pi S_2/2 = 1/16$.

The finite lattice spacing manifests itself in a lower bound to 
the real part of eigenvalues, and the finite system size studied results in oscillations
in the position of the boundary that are apparent in both the 
numerical and analytical results. A further finite-size effect, 
clear in the data but not captured in the SCBA we have presented, 
is that a finite fraction of eigenvalues are purely real: 
phenomena of this kind in random matrix problems have been 
analysed in Refs.\,(\cite{leh91}) and (\cite{efe97a}).

\section{Summary}

We have described in detail a general technique for 
calculating average spectral properties of non-Hermitian random operators.
We have applied this technique to the Fokker-Planck operator that represents particle diffusion in the presence of random advection.
In this way we obtain for the first time the
eigenvalue distribution of this operator in the complex plane.
The approach is formally exact in the weak disorder limit, 
and numerical tests show that it is remarkably
accurate even for moderately strong disorder.
Finally, we have inferred  the time-dependent effective diffusivity from the ensemble 
averaged Green's function, obtaining a novel scaling
in its transient behaviour in the strong disorder limit.

\section*{Acknowledgements}

This work was supported in part by EPSRC Grant GR/J8327.
\vspace{0.5cm}

$^\dagger${Permanent address:  Department of Theoretical and Applied Mechanics,
            Cornell University,
            Ithaca, NY 14853 }

\appendix

\section{Corrections to SCBA}
\begin{figure}[htb]
\hspace{1cm}
{\epsfig{figure=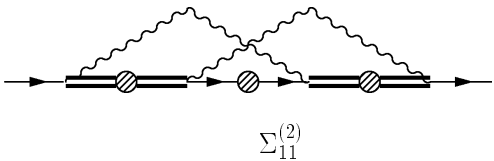,height=0.8in}}
\caption{Leading correction to the SCBA self-energy.}
\label{fig-sigma2}
\end{figure}

We demonstrate in this appendix that corrections to the SCBA 
expression for the self-energy are small in the coupling constant, $\gamma(|\omega|)$,
considering for simplicity the case $\Gamma_1 =\Gamma_2\equiv \Gamma$. 
We compare the SCBA self-energy, $\Sigma_{11}(p)$, with the 
contribution arising from the diagram illustrated in Fig.\,(\ref{fig-sigma2}),
which we denote by $\Sigma^{(2)}_{11}(p)$.
Our goal is to compute the ratio $\Sigma ^{(2)}_{11}(p)/\Sigma_{11}(p)$ in the
weak disorder limit.
We begin by calculating the correction
\bea
\Sigma ^{(2)}_{11}(p)= 
\Gamma^2 
\intl \intl d^dq_1 d^dq_2
(\p \cdot \q_2)^2 \times \nonumber \\
G_{22}(q_1)G_{11}(q_2)G_{22}(\p+\q_2-\q_1) \,.
\label{eq-a1}
\eea
First, consider the region of the complex $\omega$ plane in which the Fokker-Planck eigenvalue density, calculated within the SCBA, is zero (all $y$ for $x>0$,
and $|y|>y_B$ for $x<0$). In that region, for small $\eta$, $G_{ii}(p) \propto \eta$,
$\Sigma_{11}(p)\propto \eta$, $\Sigma^{(2)}_{11}(p)\propto \eta^3$
and the ratio has the limiting value $\Sigma ^{(2)}_{11}(p)/\Sigma_{11}(p)
= 0$ for $\eta \to 0$. 
Second, consider the complementary region ($x<0$ and $|y| < y_B$).
In this region, we use the same approach
for evaluating Eq.\,(\ref{eq-a1}) as was applied to Eqs.\,(\ref{eqq-sig1}) and (\ref{eqq-sig2})
in Sec.\,III. Substituting for $G_{ii}(p)$ in terms of $C(p)$, replacing
$C(p)$ with $I(\omega)\delta(p^2-q^2_0)$ where $Dq_0^2+x=0$, and using Eqs.\,(\ref{eq-s22}) and (\ref{eq-I}) to obtain $\Sigma_{ii}(p)$ and $I(\omega)$,
we find, for $p=q_0$ in the limit $\eta \to 0$
\be
\Sigma^{(2)}_{11}(p) = \frac{(y_B^2-y^2)^{3/2}}{\Gammab q_0^2}f
\ee
where $f$ is an angular average over directions of $\q_1, \q_2$, represented by the unit vectors
$\hat{n_1}$ and $\hat{n_2}$ 
\be
f \equiv \langle (\hat{n_1}\cdot \hat{n_2})^2 \delta(
[\hat{n}+\hat{n_2}-\hat{n_1}]^2-1)\rangle_{\hat{n_1}\hat{n_2}}
={\cal O}(1)\,.
\ee
Thus, for $|y| < y_B$, corrections to the SCBA have relative size
\be
\frac{\Sigma ^{(2)}_{11}(p)}{|\Sigma_{11}(p)|}
=\frac{y_B^2-y^2}{\Gammab q_0^2}f \sim \gamma(x)\,,
\ee
and, as advertised, are small at weak disorder.

\section{Study of discretized problem}

On a lattice, with site labels $\alpha$, $\beta$
in arbitrary dimension, the advection part of the Fokker-Planck operator,
$-\nabla \cdot \V $, is represented by a matrix with elements
\bea
A_{\alpha\beta} &=& -\frac{1}{2} u_{\alpha \beta}
\eea
for $\alpha$, $\beta$ nearest neighbor sites. It satisfies
$A_{\alpha\beta} = -A_{\beta\alpha}$ and has diagonal elements
\be
A_{\alpha\alpha} = -\frac{1}{2} \sum_{\beta}u_{\alpha\beta}
\ee
so that $\sum_{\alpha} A(\alpha,\beta) =0$, as required for conservation of particle density. Other elements are zero. We take $A_{\alpha\beta}$ to be Gaussian distributed
with zero mean, and variance
\bea
\left<A_{\alpha\beta}A_{\alpha\beta}\right> &=& \Gamma\pi^2 \hspace{0.5cm}
\left<A_{\alpha\beta}A_{\beta\alpha}\right> = -\Gamma\pi^2 \\
\left<A_{\alpha\alpha}A_{\alpha\beta}\right> &=& \Gamma\pi^2 \hspace{0.5cm}
\left<A_{\alpha\alpha}A_{\beta\alpha}\right> =-\Gamma\pi^2 \\
\left<A_{\alpha\alpha}^2\right>&=&4\Gamma\pi^2 \hspace{0.5cm}
\left<A_{\alpha\alpha}A_{\beta\beta}\right>= -\Gamma\pi^2.
\eea
for $\alpha$, $\beta$ nearest neighbor sites.
With the plane-wave basis
$\langle \r_{\alpha}|\k\rangle = N^{-1/2}e^{i\k \cdot \r_{\alpha}}$,
where $\r_{\alpha}$ is the position coordinate of site $\alpha$
on a lattice with $N$ sites in total, we need
$\langle \langle \q|A|\k\rangle \langle \k|A^T|\q\rangle \rangle$,
which takes the value
\bea
\lefteqn{\langle\left<\q|A|\k\right>\left<\k|A^T|\q\right>\rangle =}
\nonumber \\
& &\frac{1}{N^2} \sum_{\r_1 \r_2 \r_3 \r_4} 
\left<A_{\r_1 \r_2}A_{\r_3 \r_4}\right> e^{i\q \cdot (\r_1-\r_3) + i\k \cdot(\r_4-\r_2)}\nonumber \\
& & =\frac{4 \pi^2\Gamma}{N}[(1-\cos{q_x})(1+\cos{k_x}) + \nonumber \\
& &                    (1-\cos{q_y})(1+\cos{k_y})   ]\,.
\eea

\noin Similarly,
\bea
\lefteqn{\langle \left<\q|A|\k\right>\left<\k|A|\q\right> \rangle =}\nonumber \\
& & -\frac{(2\pi)^2\Gamma}{N}\left[\sin(k_x)\sin(q_x) +\sin(k_y)\sin(q_y) \right]\,.
\eea
Repeating the style of calculation described in Sec.\,III,
we find that the support of the eigenvalue density of the 
discretized Fokker-Planck operator has a boundary determined 
by 
\bea
\frac{(2\pi)^2\Gamma}{N} \sum_{q_x,q_y} \frac{(1+\cos{q_x})(2-\cos{q_x}-\cos{q_y})}
{|\omega +2D[2-\cos{q_x} -\cos{q_y}]|^2 } =1\,.
\label{eq-dis}
\eea
In the continuum limit we recover the 
result 
\bea
\Gamma \intl d^2q \frac{q^2}{|\omega+ Dq^2|^2} =1\,,
\label{eq-con}
\eea
and hence Eq.\,(\ref{eq-bdy}).

%\bibliography{/cmcl3.b/research/jwang/Paper/literature}

\begin{thebibliography}{10}
\bibitem{footnote1}
In this paper, we use the more symmetric form 
${\cal H}_0 = diag(\eta, -\eta)$ in preference to our previous
convention, ${\cal H}_0 = diag(\epsilon,-1)$ \cite{cha97}.
\end{thebibliography}

\end{document}